\documentclass[twocolumn,tighten]{aastex631}

\usepackage{cleveref}
\usepackage[dvipsnames,svgnames,x11names]{xcolor}
\renewcommand{\edit}[1]{#1}
\newcommand{\sw}[1]{\texttt{#1}}

        \newcommand{\numASCL}[0]{2104}
        \newcommand{\numJOSS}[0]{1328}
        \newcommand{\nsample}[0]{3432}

        \newcommand{\nrepos}[0]{1875}

        \newcommand{\ndevs}[0]{24753}

	\newcommand{\numgitlab}[0]{80}
	\newcommand{\numgitlabsubdomains}[0]{41}
	\newcommand{\numbitbucket}[0]{0}
	\newcommand{\percgithub}[0]{95}

\shorttitle{Scientific Software in Astronomy}
\shortauthors{Johannes Buchner}

\begin{document}

\title{High-impact Scientific Software in Astronomy and its creators}

\correspondingauthor{Johannes Buchner}
\author[0000-0003-0426-6634]{Johannes Buchner}
\email{johannes.buchner.acad@gmx.com}
\affiliation{Max Planck Institute for extraterrestrial Physics, Giessenbachstrasse, 85748 Garching, Germany}

\begin{abstract}
In the last decades, scientific software has graduated from a hidden side-product to a first-class member of the astrophysics literature. We aim to quantify the activity and impact of software development for astronomy, using a systematic survey. Starting from the Astrophysics Source Code Library and the Journal of Open Source Software, we analyse \edit{3432} public git-based \edit{scientific} software packages. Paper abstract text analysis suggests seven dominant themes: cosmology, data reduction pipelines, exoplanets, hydrodynamic simulations, radiative transfer spectra simulation, statistical inference and galaxies. We present key individual software contributors, their affiliated institutes and countries of high-impact software in astronomy \& astrophysics. We consider the number of citations to papers using the software and the number of person-days from their git repositories, as proxies for impact and complexity, respectively. We find that half of the mapped development is through US-affiliated institutes, and a large number of high-impact projects are lead by a single person. Our results indicate that there are currently over 200 people active on any given day to improve software in astronomy.
\end{abstract}





\section{Introduction}
In the last few decades, scientific software development has seen a dramatic increase in volume of code developed and professionalism of the software development \citep{asclvisionopencode}. 
This is evidenced by the now common and recommended acknowledgements of used software \citep[e.g.][]{KnapenWrite2022,ChenDataBestPractice2022}, numerous conferences dedicated to astro-informatics topics such as data pipelines and machine learning, the creation and popularity of journals for scientific software (Journal of Open \edit{Source} Software, Astronomy and Computing).
The drivers are manifold. In part, astronomical data sets have become larger, more complex and need to be processed at higher throughput, demanding systematic, computer-supported analyses \citep{Siemiginowska2019}. This includes data reduction pipelines, simulation codes, their combination with statistical data analysis, and visualisations. Potentially related to the increased complexity is that also the average number of authors of scientific publications has increased \citep{Kerzendorf2019}. A further driver may also be the increased ease of writing and publishing high-level software and the available computing power due to Moore's law making a wider class of simulations and data analyses feasible (e.g., most recently, deep learning).

The explosion of scientific software however still is poorly quantified. Some individual high-impact software packages, such as astropy, have garnered an impressive number of citations, and their impact and \edit{sustainability} was analysed by \cite{SunSustainAstropy2024}. However, a full exploration of the landscape is still missing. This work aims to bring the work and authors behind high-impact scientific software in astronomy into the spotlight.

\section{Data}
We attempt a quantitative investigation of scientific software development and its impact to astronomy research.
To this end, we first define the sample of scientific software considered for this study, and operationally define impact and development activity.
Our data analysis is open source and reproducible\footnote{available at \url{https://github.com/deepthought-initiative/architects-of-modern-astronomy}}.

\subsection{Sample}
For this study, we use two parent samples. The first is the Astrophysics Source Code Library \citep[ASCL][]{1999AAS...194.4408N,2020ASPC..522..731A}. Each ASCL entry links to associated paper(s) and to the code. The paper links are typically to the Astrophysics Data System (ADS) Bibliographic Services, which we use to trace citations to the code. The code links either lead directly to code repositories (such as GitHub, GitLab, BitBucket, Sourceforge, Hepforge or similar), or to websites, in which case we parse the website for links to code repositories. In case of multiple code links, we use the first reachable git repository.
The second sample is based on papers published in the Journal of Open Source Software (JOSS). We obtain this sample from ADS and require at least 2 citations. \edit{This is to focus on astronomy-related JOSS papers.} The ADS link leads to JOSS, which links to the code repository. 

We deduplicate entries by git repository link.
We only use entries with a publicly reachable git repository. \edit{This is a important limitation of the sample used in this work.}

The sample contains \nsample{} entries (\numJOSS{} from JOSS, \numASCL{} from ASCL), fetched in November 2025. 
In \percgithub{}\% of cases, the repository is hosted on GitHub. However, the sample also includes \numgitlab{} entries from GitLab, \numbitbucket{} entries from BitBucket and \numgitlabsubdomains{} GitLab instances hosted on local subdomains (such as gitlab.mpcdf.mpg.de and gitlab.in2p3.fr).

\subsection{Impact}
Defining how important a piece of software (or any work) is, is subjective. Any definition will have \edit{shortcomings}.
For investigating the impact of thousands of codes, we require a definition that can be algorithmically determined. To measure the scientific impact, we can use the astronomical literature to see how many papers used the software, and how impactful those papers were.
The impact of scientific paper is frequently measured in terms of number of citations to the paper. 
We could therefore count the number of citations to the software.
However, here, we are rather interested in whether papers using the software were impactful themselves (a second-order impact). Therefore, we count the number of citations of astronomy papers ('collection:astronomy' in ADS) citing software $S$:
\[
\mathrm{Impact}(S) = \sum_{p\in\mathrm{astrocitations(S)}} |\mathrm{citations}(p)|
\]
Software that has not been cited, according to this definition, had no impact. In case of multiple associated papers (in ASCL), papers that cite any of them are identified.
We recognize the shortcoming of this definition: Sub-fields differ in their citation habits, and citations are an imperfect tracer of scientific value.

\begin{figure}
    \centering
    \includegraphics[width=\columnwidth]{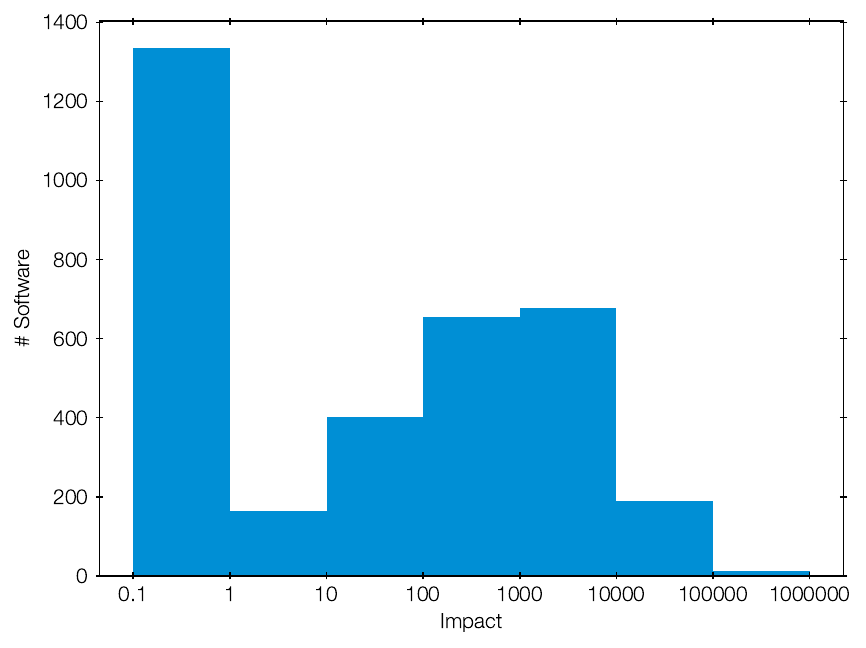}
    \caption{Sample distribution of the impact of scientific software. Impact is the total number of citations to papers using the software.}
    \label{fig:impact}
\end{figure}

\Cref{fig:impact} shows a bi-modal distribution of impact values. A large number of software packages have no citations. This may be because they are not cited in astronomy. A mode is apparent near 1000 citations, and some packages accumulate 100000 citations or more.
A noteworthy observation here is that it is not uncommon for software papers to impact thousands of scientific papers.

\subsection{Creators \& Effort}
Next, we identify the people who create the software.
Co-authorship on the associated paper does not reliably reflect software contributions. To measure software contributions, we retrieve the contributor history from the git repository. Typically, each git commit is an atomic change to the software. However, the frequency and meaning of how much change warrants a git commit varies across developers. Furthermore, it is difficult to judge the importance of a repository contribution.
To quantify the contribution of each person, a definition is needed that can be applied across such variations.

\edit{As an effort metric, we count the number of days on which an author made a commit. By counting the number of active days}, we attempt to trace an author spending (some of) the day on the software concentration and thinking.
We obtain each git commit timestamps, deduplicate subsequent timestamps within 86400 seconds into the same day, and then count the number of unique days.
By this time-zone-independent measure, an author making 1 commit every day would be counted more than an author making 355 commits in one day once every year, however, this is an artificial example. In most projects, the commit history is dominated by one or two developers, with a long list of occasional contributors who fix one bug or add a piece of documentation. We keep only major contributors, which we define as having spent at least 10 per cent of the number of days of the most active contributor.

We identify unique authors by their names from the git log. Git author names may slightly differ from the author names used in the software paper. We use partial matching (firstname and lastname, lastname and firstname, lastname and initial, firstname as git user only, etc) to associate git contributions to the paper author where possible. This also helps with deduplication of git contributor names.

\edit{Finally, in some of our results we show the multiplication of software impact times effort, aggregated by author, software, and affiliation (see next section). The units of this metric are citations$\times$person-days. This metric is intended to grow with both usefulness and project difficulty.}

\subsection{Supporting institutions}
For academic institutions, it may also be interesting to see the impact of their contributions. 
For the community that benefits from open source software, it is relevant to know the supporters and funders of open source development. 
In a future work, we may aim to extract funding information from the full text of each paper (for example from the acknowledgement section). However, this information is not easily available.

We identify institutions using paper affiliations. Above, we already mapped git code authors to paper authors, where possible.
A author identity is then available through the associated paper giving author names when an Open Researcher and Contributor ID (ORCID) is set. 
In that case, we use the ORCID to fetch the author publication history. From the papers, for each calendar year, we obtain a set of affiliations used. We only use papers within the date range from the first to the last git commit in the history for each person, with one extra year of margin. Finally, we combine the set of paper affiliations.
We heuristically deduplicate affiliations with slightly different spelling (e.g., English/German/Spanish, "Department of Astronomy, University of X" vs "University of X").
Each affiliation found is assigned the (same) contribution of the author. This has the shortcoming of double counting a software package for multiple institutions. Therefore care needs to be taken when summing several institutions. Despite this, institutions can be compared among each other.

\begin{figure*}
    \centering
    \includegraphics[width=\textwidth]{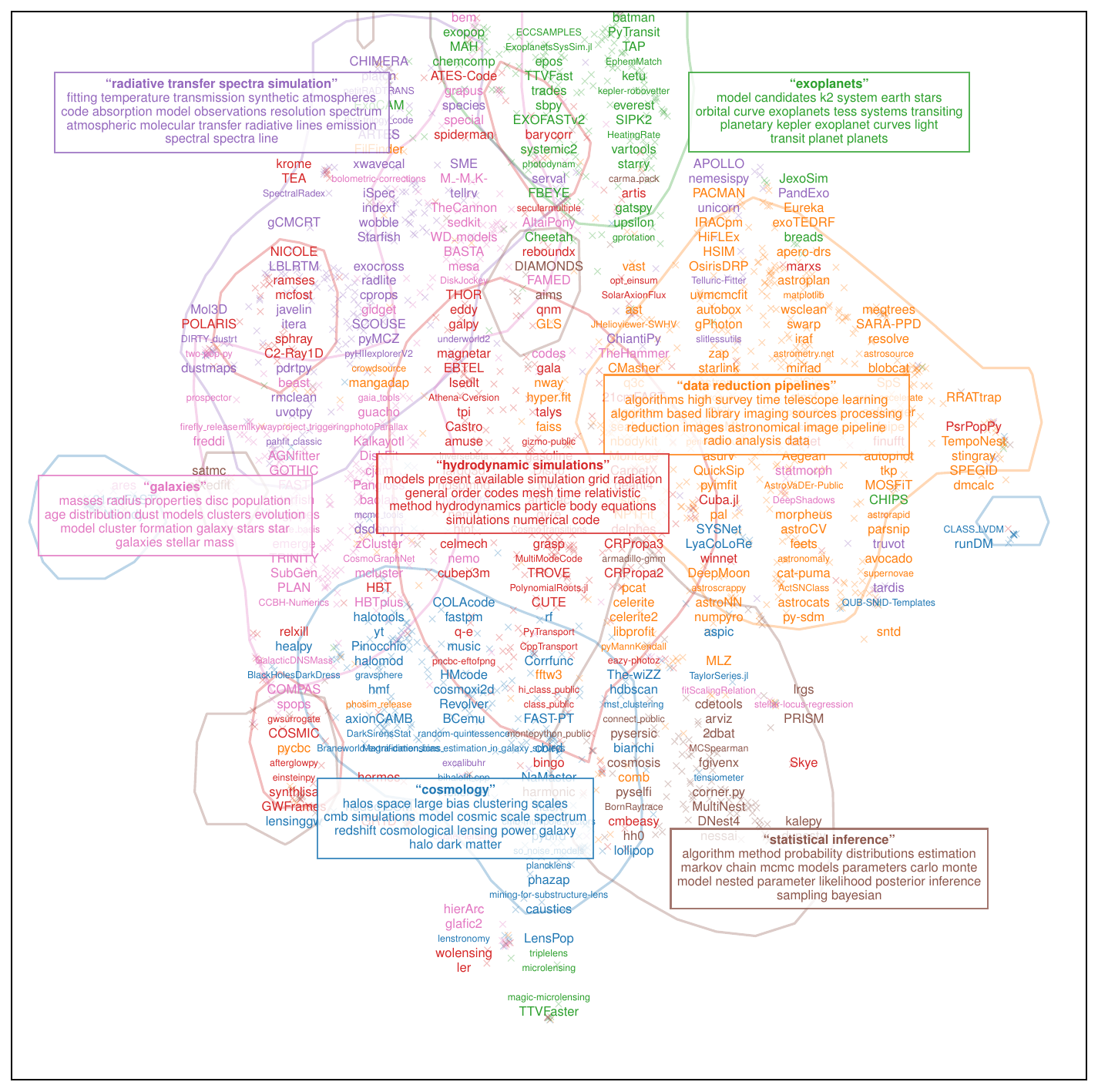}
    \caption{Landscape of astronomical software.
    \edit{Coloured contours and rectangles identify the seven principal themes from our bag-of-words analysis. In the shown 2-dimensional projection, representative software packages are named, and color-coded by most important theme. Each code is marked by a small cross.}
    }
    \label{fig:landscape}
\end{figure*}

\subsection{Data quality issues}
\label{sec:dataqual}
\edit{Due to the automated analysis, the assembled data has noteworthy quality issues. This includes (1) the incomplete work audit when repositories are imported from another version control system, (2) the incomplete mapping of git author to paper author and their affiliated institute and country, and (3) the double-counting in case of multiple affiliations. 
Here we point out some egregious problems. \sw{IRAF} is attributed to developers who contributed after 2017, when the Github repository was started, instead of the National Optical Astronomy Observatories (NOAO), where it was developed and maintained until 2013, including by Doug Tody, Lindsey Davis and Frank Valdes\footnote{\url{https://adass.org/softwareprize.html}}. One referee points out that \sw{starlink} is attributed to affiliations by Legacy Survey of Space and Time (LSST), Cornell University, University of Washington and observatories in Hawaii, which is likely due to the multiple affiliations of a currently active developer. However, \sw{starlink} was originally a UK software project operating from 1980-2005 with very different version control operation. The attribution to developers and institutes here is likely unreliable and due to later maintainance. There are likely more data issues that have not yet been noticed. Handling more data cleaning is left for future work. Contributions to the reproducible analysis code repository are welcome. For now, the data are cautiously interpreted as they are.}

\subsection{Principal themes}\label{sec:themes}
We discover the principal themes of astronomical software by applying data-driven text analysis to associated paper titles and abstracts. We adopt a standard a bag-of-words framework and use standard techniques implemented in \texttt{scikit-learn}. We first tokenize each text into words, excluding English stop words, and identify the relative over-abundance of each word \citep[term frequency–inverse document frequency, TF-IDF][]{robertson2004tfidf}. Next, over this non-negative high-dimensional feature space,  we identify 7 principal components using non-negative matrix factorization \citep{lee1999NMF}. Increasing the number to 8 or 9 gives thematic repetitions. For each principal component, the 20 most important features (words) are identified. Next, each software is classified into these themes, by the theme corresponding to the highest principal parameter. 

Finally, the topical distribution of softwares is visualised in a two-dimensional space. Here we adopt Uniform Manifold Approximation and Projection \citep{McInnes2018UMAP} using TF-IDF cosine distances, not relying on the NMF results above.

\begin{figure*}
    \centering
    \includegraphics[width=\textwidth]{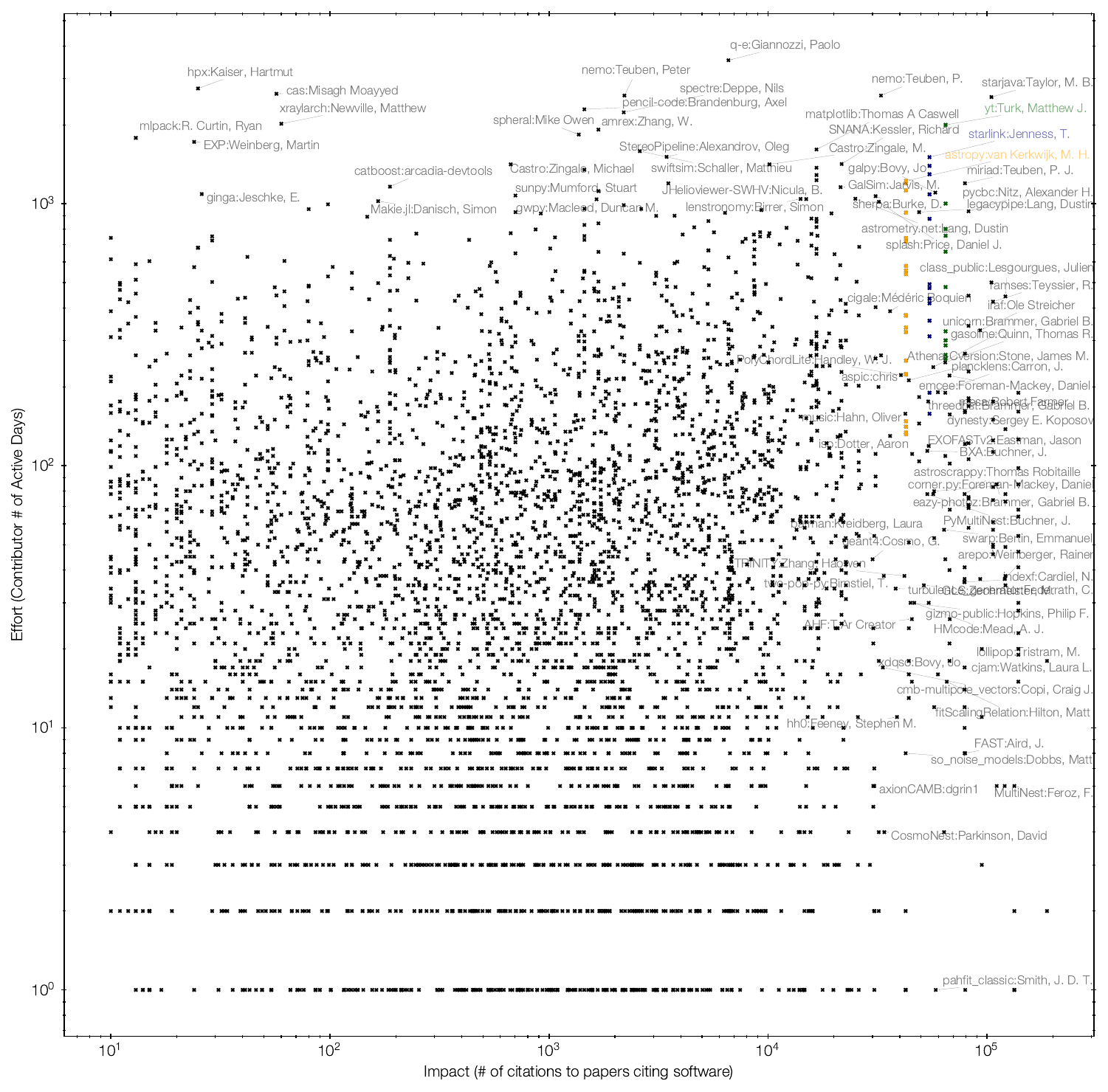}
    \caption{Comparing project impact and contribution per author.
    Some individual points are annotated with project name and most-contributing author. \sw{astropy} is in orange, \sw{starlink} in blue, \sw{yt} in green.
    }
    \label{fig:impact_contributions}
\end{figure*}

\begin{figure*}
    \centering
    \includegraphics[width=\textwidth]{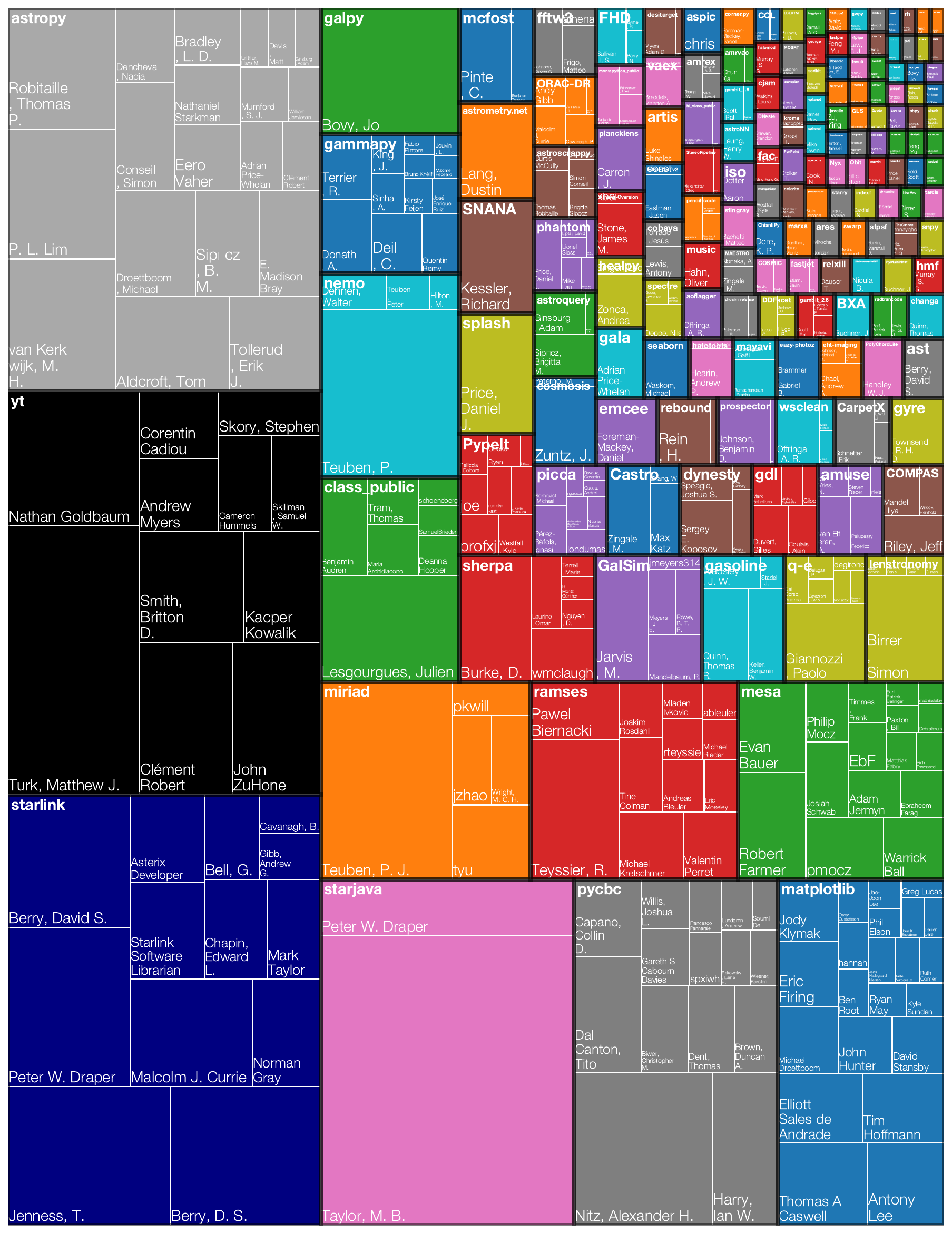}
    \caption{A tree map visualisation of astronomical software. Each white rectangle identifies one author. The rectangle is proportional to scientific impact multiplied by number of days spent. Each black rectangle is one software, with the name at the top left. The smallest are excluded in this figure.
    }
    \label{fig:withouttiny}
\end{figure*}
\begin{figure*}
    \centering
    \includegraphics[width=\textwidth]{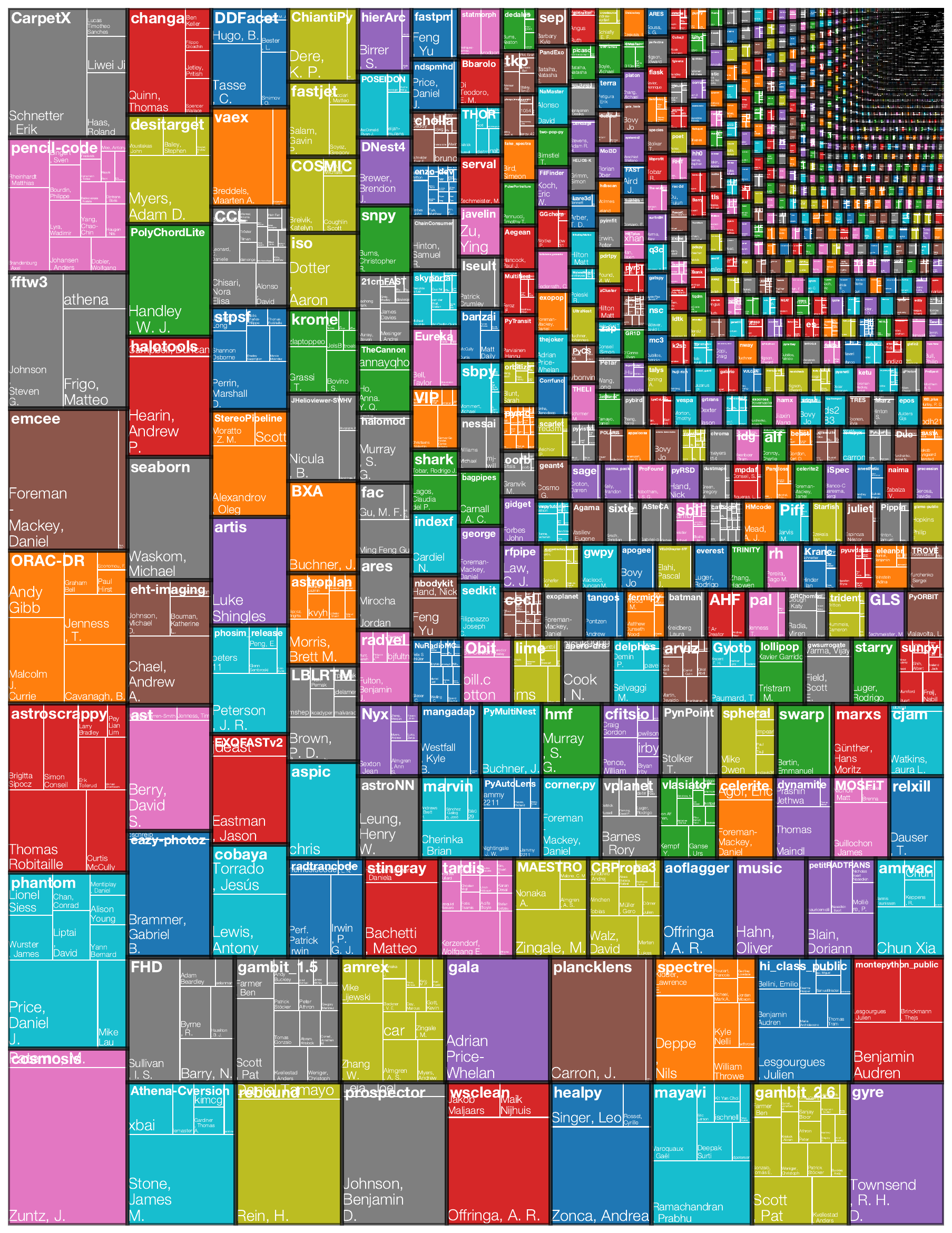}
    \caption{Same as \cref{fig:withouttiny}, but removing the top 30 packages, and showing smaller ones.
    }
    \label{fig:withoutmajor}
\end{figure*}

\section{Results}
With the time effort and impact measured, we visualise the landscape of astronomical open source software.
The seven principal themes are listed by their most important 20 words in the rectangles of \cref{fig:landscape}. They can be roughly summarized as: \textbf{cosmology, data reduction pipelines, exoplanets, hydrodynamic simulations, radiative transfer spectra simulation, statistical inference and galaxies}. A map of software is visualised in \cref{fig:landscape}, with each software marked as a cross, colored by theme. For each theme, we identify 68\% contours with kernel density estimation. Using a grid over the 2-dimensional space, we label the highest-impact software in each cell.

Figure~\ref{fig:impact_contributions} compares the time effort put into software with the literature impact. These include general-purpose tools such as \sw{starlink}, \sw{starjava} (including \sw{topcat}), \sw{astropy}, \sw{yt} and \sw{matplotlib}. There is no strong apparent correlation between days and number of citations. However, especially in older software, the git history may not include the full software history, and the paper associated with the software via ASCL or JOSS may not be the one receiving most citations. One of the packages with high impact and effort is \sw{astropy} (orange dots in Figure~\ref{fig:impact_contributions}) and \sw{starlink} (blue dots). These show a large number of contributors.

For the rest of this work, we focus on the multiplication of impact and effort. This is assumed to trace the complexity of software and importance of software together, and indicates major fruitful software development efforts. \edit{We only consider works with more than 10 citations to its paper(s). This leaves \nrepos{} git repositories.}

Figure~\ref{fig:withouttiny} shows the dominant software packages, measured by the multiplication of impact and developer active days.
These include general-purpose tools such as \sw{starlink}, \sw{starjava} (including \sw{topcat}), \sw{astropy}, \sw{yt} and \sw{matplotlib}.
Figure~\ref{fig:withoutmajor} is the same visualisation after but zooming in beyond the top 30 software packages.

\begin{figure*}
    \centering
    \includegraphics[width=\textwidth]{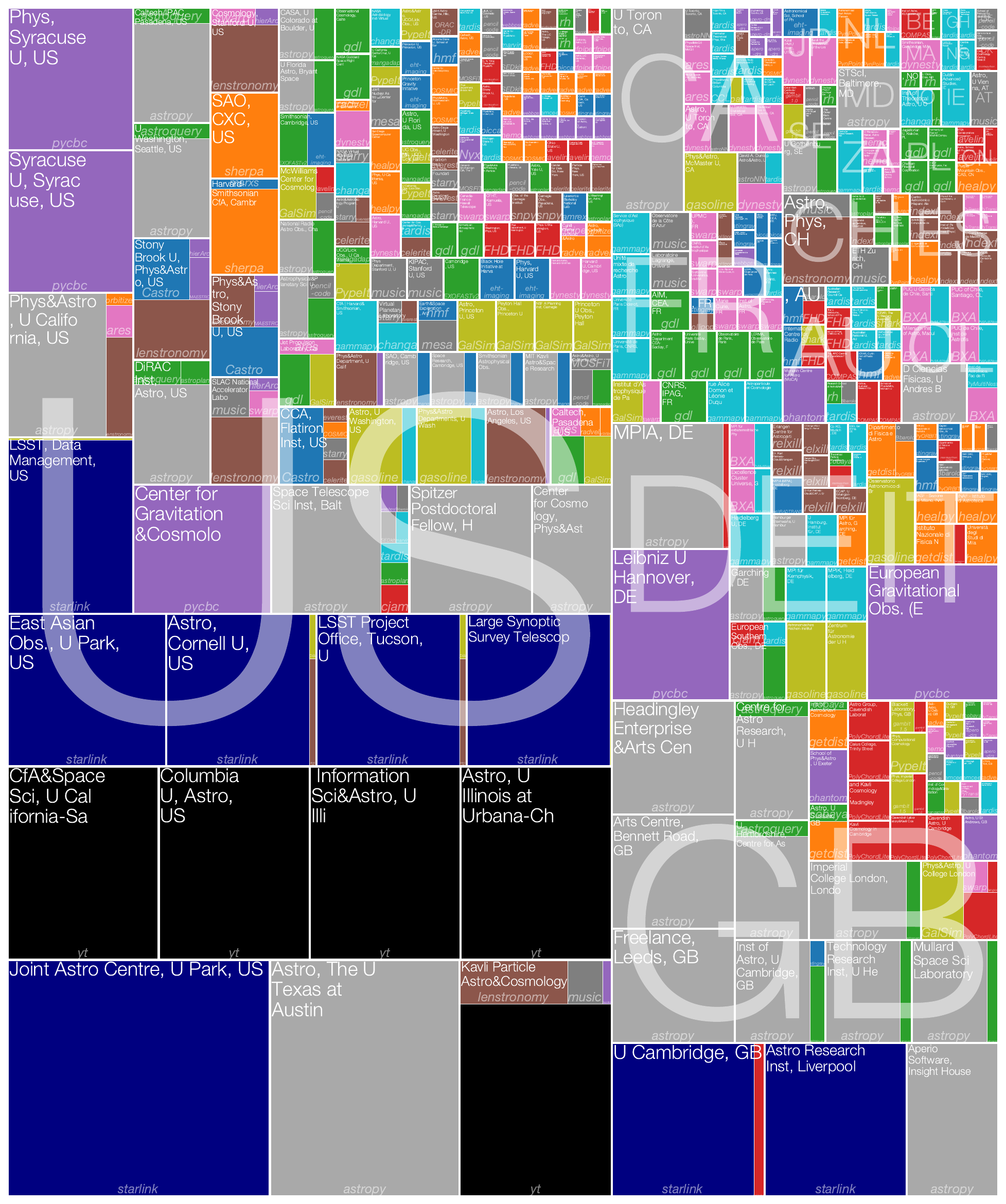}
    \caption{A tree map visualisation of astronomical software. Each white rectangle identifies one institution (from affiliation). The rectangle is proportional to scientific impact multiplied by number of programming days spent. \edit{Contributions from authors with multiple affiliations are counted as many times.}
    Countries are identified with large gray letters. Each color corresponds to a project.
    }
    \label{fig:countries}
\end{figure*}

\Cref{fig:countries} presents a visualisation of the supporting institutions. These are split by countries (large gray letters), with the left rectangle being US, and the right including rectangles for Great Britain, Germany, Canada, France, Italy, South Africa, Chile, Australia, Switzerland etc.
Colors indicate projects. For example, \sw{yt} (black rectangles) is developed in Illinois, Columbia University and University of California. \edit{\sw{starlink} is stated as developed by Legacy Survey of Space and Time (LSST), Cornell University and University of Washington and observatories in Hawaii. However, one referee points out that \sw{starlink} was originally a UK software project operating from 1980-2005 with very different version control operation. The attribution to developers and institutes here is likely unreliable}.
\sw{astropy} (in grey) development is notably spread over many countries and institutions.

\section{Discussion and Conclusion}
This work has made a systematic survey of software developed and used in astronomy. 
Previously, the depsy\footnote{http://depsy.org/} project \citep{singh2016unsung} started from python and R package databases, and searched the content of scientific papers for citations to each software. Focusing on these two languages, they attempt to build a dependency tree of the software stack, so that software that is being implicitly relied upon receives credit on its impact. However, the parsing required is complex and limited \citep{singh2016unsung}. In this work, we follow a language-agnostic approach, and focusing on astronomy, use the reliable ADS to query software-associated papers.

This is the first systematic study of astronomical software to connect a work audit trail to the impact to the scientific discourse and funding bodies. For practical reasons, we have limited ourselves to open source git repositories to access the git log. We query ADS to measure second-order citation impact and to identify affiliated institutions and countries. We have found over 1500 highly cited software packages, with development effort ranging from person-days to person-years. 
\edit{There is no obvious correlation between impact and number of person-days.}
Besides a few large, well-known projects, there is a wide array of impactful astronomy software (see \cref{fig:withouttiny,fig:withoutmajor}), related to visualization, data reduction, statistical analysis, simulation, and other functionality.

\edit{Due to the automated analysis, the assembled data has noteworthy quality issues, as discussed in section \ref{sec:dataqual}. The focus on public git repositories introduces severe incompleteness to the traced development activity. Older software is either missed, or if imported into git, can lose the work trail and affiliations, skewing the credit assigned for high-impact work. Publications may undercite research software, in part due to missing guidance by astronomy and physics journals on how to acknowledge software.}

\Cref{fig:withouttiny,fig:withoutmajor} recognises the individual contributors that enable great science, traced by highly cited papers using that software. Top software packages appear preferentially developed by a substantial number of contributors who contribute nearly equally (see \cref{fig:withouttiny}). \edit{However, in both the high and low-impact regime, we identify packages developed nearly equally by a group of developers, and packages where development is dominated by a single person (see \cref{fig:withouttiny}).}
The challenges in sustaining software development have been discussed elsewhere, e.g., in \cite{Siemiginowska2019} and \cite{SunSustainAstropy2024}. 

We can also estimate a lower bound on the number of active developers. We find a total of \ndevs{} unique email addresses mentioned in git logs.
Counting the number on each calendar day, we find an increase in unique email addresses over the years. Taking a median over 120-day windows to remove holiday effects, \edit{we find a median of \ndevs{} active developers in mid-2020}. Past then, the data is likely incomplete as software packages are not yet published. From this, we estimate that there are currently at least \ndevs{} people working actively on any given day on creating and improving astronomy software.

The challenges in funding the creation of complex software have also been pointed out \citep[e.g.][]{Siemiginowska2019}. \Cref{fig:countries} recognises the countries and institutions that presumably support high-impact software, and the careers of their developers.

Future work could more closely investigate funding bodies or the career of developers. However, this would require either a funding/career database or parsing paper/CV contents. Refined metrics for quantifying the development activity and its scientific impact could also be developed.

\section*{Acknowledgements}
This research has made use of NASA’s Astrophysics Data System and its python interface\footnote{\url{https://github.com/andycasey/ads/}}. For analysis, we used \sw{scikit-learn} \citep{scikit-learn}, \sw{corner} \citep{FMD2016corner}, \sw{matplotlib} \citep{Hunter2007} and the treemap from \sw{matplotlib-extra}\footnote{\url{https://github.com/chenyulue/matplotlib-extra}}. JB thanks Wolfgang Kerzendorf for insightful conversations about this manuscript.

\section*{Code and Data availability}
Analysis code and data are available at \url{https://github.com/deepthought-initiative/architects-of-modern-astronomy}.

\bibliography{biblio}
\bibliographystyle{aasjournal}
\end{document}